\documentclass[letterpaper,preprint,superscriptaddress,floatfix,showpacs]{revtex4}
\usepackage[latin1]{inputenc}
\usepackage{bm}
\usepackage{color}
\usepackage{multirow,amssymb,amsbsy,amsmath}
\usepackage{graphicx}
\usepackage{verbatim}
\usepackage{upgreek}

\makeatletter
\usepackage{pifont}
\makeatother

\begin{document}

\title{Experimental certification of ensembles of high-dimensional quantum states with independent quantum devices}

\author{Yong-Nan Sun}
\affiliation{School of Physics, Hangzhou Normal University, 311121 Hangzhou, Zhejiang, China}

\author{Meng-Yun Ma}
\affiliation{School of Physics, Hangzhou Normal University, 311121 Hangzhou, Zhejiang, China}

\author{Qi-Ping Su$\footnote{email:sqp@hznu.edu.cn}$}
\affiliation{School of Physics, Hangzhou Normal University, 311121 Hangzhou, Zhejiang, China}

\author{Zhe Sun}
\affiliation{School of Physics, Hangzhou Normal University, 311121 Hangzhou, Zhejiang, China}

\author{Chui-Ping Yang$\footnote{email:yangcp@hznu.edu.cn}$}
\affiliation{School of Physics, Hangzhou Normal University, 311121 Hangzhou, Zhejiang, China}
\affiliation{Quantum Computing Center, RIKEN, Wakoshi, Saitama, 351-0198, Japan}

\author{Franco Nori$\footnote{email:fnori@riken.jp}$}
\affiliation{Quantum Computing Center, RIKEN, Wakoshi, Saitama, 351-0198, Japan}
\affiliation{Physics Department, The University of Michigan, Ann Arbor, MI 48109-1040, USA}


\begin{abstract}

When increasing the dimensionality of quantum systems, high-dimensional quantum state certification becomes important in quantum information science and technology. However, how to certify ensembles of high-dimensional quantum states in a black-box scenario remains a challenging task. In this work, we report an experimental test of certifying ensembles of high-dimensional quantum states based on prepare-and-measure experiments with \textit{independent devices}, where the state preparation device and the measurement device have no shared randomness. In our experiment, the prepared quantum states are high-dimensional orbital angular momentum states of single photons, and both the preparation fidelity and the measurement fidelity are about 99.0$\%$ for the six-dimensional quantum states. We also measure the crosstalk matrices and calculate the similarity parameter for up to ten dimensions. We not only experimentally certify the ensemble of high-dimensional quantum states in a semi-device-independent manner, but also experimentally investigate the effect of atmospheric turbulent noise on high-dimensional quantum state certification. Our experimental results clearly show that the certification of high-dimensional quantum states can still be achieved even under the influence of atmospheric turbulent noise. Our findings have potential implications in quantum certification and quantum random number generation.
\end{abstract}

\pacs{42.50.Dv}

\maketitle

With the development of quantum technologies, the precise control of complex quantum systems or devices offers potential applications in fields such as quantum communication \cite{Briegel1998}, computation \cite{Reichardt2013}, cryptography \cite{Gisin2002}, and sensing \cite{Giovannetti2011}. To precisely and efficiently characterize quantum states, various techniques have been developed for quantum state characterization, each tailored and optimized for specific scenarios, such as quantum state tomography \cite{Hradil1997,Matteo2004}. Although quantum state tomography is powerful and applicable, it relies on the assumption about the inner workings of quantum devices; namely, a well-calibrated measurement device is required to certify the intended quantum state. However, realistic quantum devices are always imperfect: they suffer from unavoidable noise and decoherence; especially in some cases, they are not trusted. In order to ensure that a fabricated device can be applied to effectively perform the intended quantum operations and generate the necessary quantum states, it must undergo a state certification procedure. Hence, how to efficiently certify and characterize quantum states is crucial in quantum information processing.

The requirement of precise controls in quantum state tomography can be released through the utilization of advanced certification methods: device-independent (DI) or semi-device-independent (SDI) certification. In the fully DI approach, quantum devices are prepared as a black box, and quantum states or measurements are certified only by measurement results, without making assumptions about detailed functioning of the devices used in experiments \cite{supis2020}. Moreover, DI certification of entangled states or measurements can be realized through the violation of the Bell inequality \cite{chen2021,zhao2023,chen2016}. Although a number of experiments have been implemented in a DI manner \cite{Smania2020, Zhang2018}, this elegant approach faces some obstacles in experiments. Especially, it requires the maximal violation of certain Bell inequalities, which is very vulnerable to experimental imperfections.

Alternatively, with an additional assumption on the dimension of the Hilbert space, or extra assumptions, such as mean energy or minimal overlap \cite{Brask2017, Himbeeck2017, Chaves2015}, the SDI approach can be applied to certify quantum states and measurements in prepare-and-measurement experiments, which are more feasible to implement and tolerant of realistic experimental imperfections \cite{Hendrych2012, Ahrens2012, Tavakoli2015, DAmbrosio2014, Lunghi2015, Martinez2018, Sun2020, Biagi2021}. To date, SDI certification has garnered significant attention, resulting in the development of various certification methods for qubit states \cite{Tavakoli2018, li2023,Tavakoli12024}, mutually unbiased bases measurements and symmetric informationally complete measurements \cite{Farkas2019, Tavakoli2019,Gabor2024}, nonprojective qubit and qudit measurements \cite{Tavakoli2020, Mironowicz2019, Fan2022, Martlnez2023}, and qubit quantum instruments \cite{Mohan2019, Miklin2020}.

Previous works on quantum state and measurement certification using the SDI approach have adopted the models in which quantum devices can be classically correlated in a stochastic and unknown manner. As discussed in \cite{Armin2020, Bowles2014}, it is more natural to assume that all devices are not correlated in a practical setup, i.e., they are independent. Previous works on quantum state certification using the SDI approach are limited to 2D quantum states \cite{Tavakoli2018, li2023}, but \textit{experimentally realizing SDI certification of high-dimensional quantum states remains a challenging task}.

Motivated by the above, in this work we experimentally certify for the first time ensembles of high-dimensional quantum states with independent devices. We can certify ensembles of high-dimensional states corresponding to quantum $t$ designs, which are based on a randomized version of unambiguous state discrimination in a SDI manner. A quantum $t$ design \cite{Delsarte1977, Renes2004} is a set of $d$-dimensional quantum states, which has the property that the average of any $t$-degree polynomial over the set of $d$D states equals to the average taken over all pure states. The ensembles of quantum states corresponding to quantum $t$ designs have broad applications in quantum information science, including quantum tomography \cite{Hayashi2005, Scott2008}, quantum key distribution \cite{Renes20042}, entropic uncertainty relations \cite{Ketterer2020}, and entanglement detection \cite{Liu2018,Bae2019}.

In our work, the prepared quantum states are high-dimensional orbital angular momentum (OAM) states of single photons. We further investigate quantum state certification under discrimination errors. We study the effects of atmospheric turbulence on high-dimensional OAM states. The atmospheric turbulence is simulated by a single phase screen based on the Kolmogorov theory of turbulence. Our experimental results show that we can still achieve the certification of high-dimensional quantum states even under the influence of atmospheric turbulence noise.

\begin{figure}[tbph]
\centering
\includegraphics[width=4in]{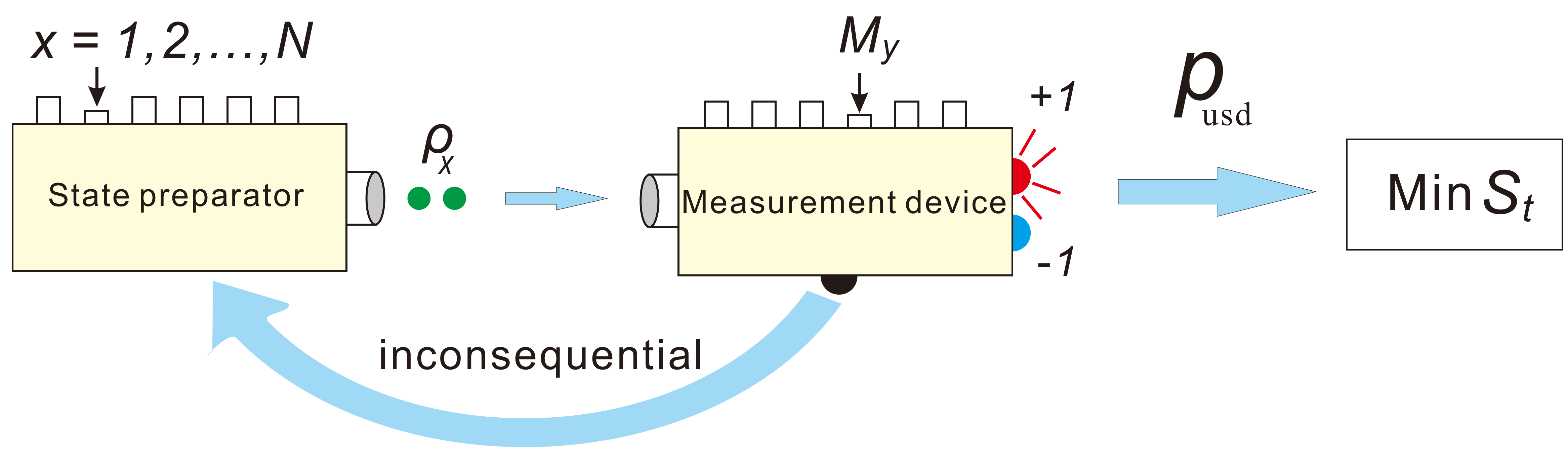}
\caption{Randomized unambiguous state discrimination.}
\end{figure}

We consider two devices (Fig.~1), the state preparation device and the measurement device. The box features $N$ buttons that label the prepared states. When pressing button $x \in \{1,...,N\}$, the box emits a particle in the state $\rho_{x}$. For the state certification, the emitted particles are sent to the measurement device. When button $y$ is pressed, the device performs measurement $M_{y}$ on the incoming particle and the corresponding measurement outcome is $b$.

The certification of quantum states and measurements is realized with randomized unambiguous state discrimination (USD). In randomized USD, as shown in Fig. 1, Alice has two random inputs $x_{1}$, $x_{2}\in \{1,...,N\}$ ($x_{1}<x_{2}$). In each experiment, Alice prepares the corresponding quantum states $\vert \psi_{x_{1}}\rangle$, $\vert \psi_{x_{2}}\rangle$ based on the two random inputs $x_{1}$, $x_{2}$. Then Alice sends the quantum states $\vert \psi_{x_{1}}\rangle$, $\vert \psi_{x_{2}}\rangle$ to Bob. Bob has one random input  $y \in \{1,...,N\}$ and performs the corresponding measurement $M_{y}=\{M_{b|y}\}_{b}$. The measurement yields three possible outcomes $b\in \{-1, 1, \perp \}$, where $\perp $ denotes the inconclusive outcome. Bob's task is to unambiguously discriminate between Alice's two states $\{\vert \psi_{x_{1}}\rangle,\vert \psi_{x_{2}}\rangle\}$, otherwise the round is inconsequential.

In the randomized USD, Alice's device does not require precise characterization, but is assumed to produce quantum states with a known Hilbert space dimension. The performance in the randomized USD relies on the outcomes of all individual USD tasks. The rate of inconclusive rounds is consistently $1-p^{x_{1},x_{2}}_{\mathrm{usd}}$,
\begin{equation}
p^{x_{1},x_{2}}_{\mathrm{usd}}= \dfrac{1}{2}[p(1|\psi_{x_{1}})+p(2\vert \psi_{x_{2}})].
\end{equation}

Subsequently, we consider all the inconclusive events accumulated across the individual $\mathrm{USD}$ tasks \cite{Armin2020}:
\begin{equation}
S_{t}\equiv \sum_{x_{1}, x_{2}=1}^{N}(1-p^{x_{1},x_{2}}_{\mathrm{usd}})^{2t},
\end{equation}
where $x_{1}<x_{2}$ and the integer $t\geq1$ is the order of the inconclusive events. Hence, randomized USD is characterized by three parameters: the dimension $d$, the ensemble size $N$, and the order $t$. Striving for optimal USD performance across all values of $y$ implies that Alice and Bob are working towards minimizing the value of $S_{t}$. By exploiting the independence of the devices, the minimal value of $S_{t}$ is attained when utilizing pure states \cite{Armin2020}. Hence, in order to evaluate Eq.~(2), we use the initial state $\rho_{x}=\vert \psi_{x}\rangle \langle \psi_{x}\vert$. Therefore, for all states produced by Alice and input $y$, the minimum value of $S_{t}$ \cite{Armin2020} is 
\begin{equation}
\rm \min\limits_{\text{quantum}}\mathit{S_{t}}=-\dfrac{\mathit{N}}{2}+\dfrac{1}{2}min\sum_{x_{1},x_{2}}|\langle \psi_{x_{1}} \vert \psi_{x_{2}}\rangle|^{2\mathit{t}}.
\end{equation}

Utilizing quantum $t$ designs, we can derive that the optimal quantum implementation of randomized USD obeys
\begin{equation}
\rm \min\limits_{\text{quantum}}\mathit{S_{t}}\geq \dfrac{1}{2}\left[\dfrac{\mathit{N}^{2}\mathit{t}!(\mathit{d}-1)!}{(\mathit{t}+\mathit{d}-1)!}-\mathit{N} \right].
\end{equation}
The bound can be achieved if and only if Alice's states constitute a quantum $t$ design of dimension $d$, which consists of $N$ states. Therefore, this concludes the certification. In our experiment, we certify a family of quantum $t$ designs, which correspond to $t=2$ and $N=d^{2}$ for any $d\geq2$. These special states, known as symmetric informationally complete (SIC) \cite{Renes2004} states, can be also used in entanglement detection \cite{Shang2018} and unambiguous state discrimination of nonorthogonal states \cite{Dieks1988, Peres1988}.

In our work, the prepared quantum states are high-dimensional OAM states of single photons. For different dimensions, we first randomly prepare high-dimensional quantum states for each dimension, then perform measurements. Next, the measurement probabilities are substituted into Eq. (2) for computation, and the computed results are compared with the theoretical values. If the experimental results align with the theoretical predictions, it indicates that the states we prepared are symmetric informationally complete quantum states.

\begin{figure}[tbph]
\includegraphics[width=5.5in]{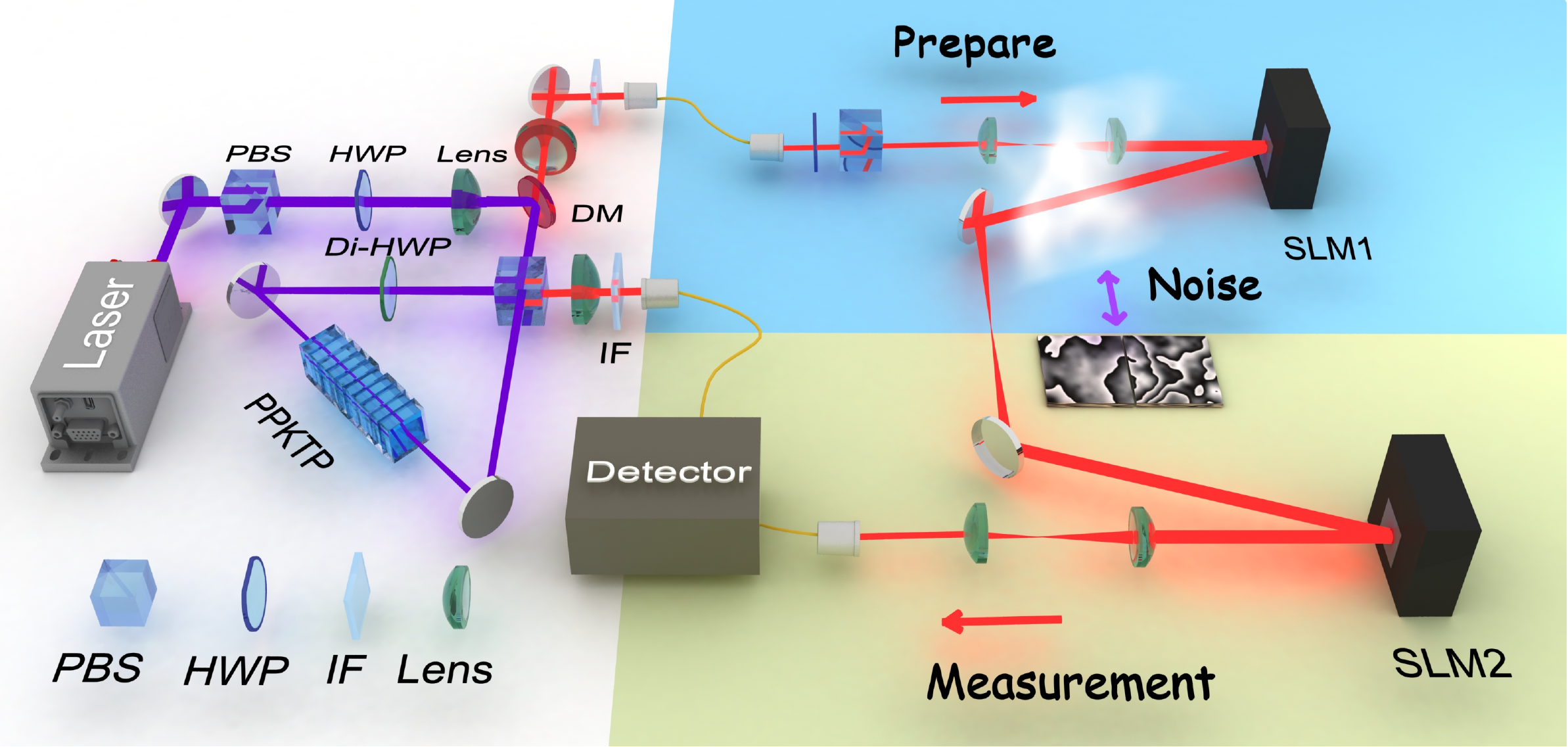}
\caption{Experimental setup for certifying ensembles of states with independent devices. The heralding photon, produced by spontaneous parametric down-conversion, is directly sent to a detector which acts as a trigger. The signal photon is projected on the fundamental Gaussian state by means of a single mode fiber. In order to make full use of the SLM, we use two lenses to expand the laser beam. To avoid the Gouy phase shift effect, an imaging $4f$ system is implemented between the screens of the two spatial light modulators. PBS stands for polarization beam splitter; Di for dichroic; DM for dichroic mirror; and IF for interfering filter.}
\end{figure}

As shown in Fig.~2, a single-mode laser with a wavelength of 405 nm is used to pump the periodically poled $\rm KTiOPO_{4}$ (PPKTP) nonlinear crystal, which is placed into a phase-stable Sagnac interferometer (SI) to generate photon pairs at 810 nm. A polarized beam splitter (PBS) followed by a half-wave plate (HWP) is used to control the polarization mode of the pump beam. Lenses situated before and after the SI are used to focus the pump light and collimate the photon pairs, respectively. To remove the pump beam light, a dichroic mirror (DM) is used in our experiment. After passing through the interference filter (IF, $\Delta \lambda$ = 3 nm, $\lambda$ = 810 nm), the photon pairs, which are generated in the spontaneous parametric down conversion (SPDC) process, are coupled into single-mode fibers separately. When the power of the pump beam is 2.67 mW, the total coincidence count collected by the SPD is about $3 \times 10^{4}$ per second. The coincidence time window is 1 ns and the data acquisition time is 3 s.

The OAM plays important roles in quantum information and quantum foundations \cite{37,Bliokh2017,39,Bliokh2012}. With OAM, we can generate high-dimension quantum states. As shown in Fig. 2, the signal photon is projected onto the Gaussian state through a single-mode fiber. After spatial filtering and expanding with two lenses, the OAM states are manipulated with the first spatial light modulator (SLM) in order to prepare the desired superposition states. The SLMs used in our experiment are phase-only, which means we can directly control the phase of the light field through our SLM, but the amplitude of the light field can only be controlled indirectly. Various effective methods have been developed to encode high-dimensional single-photon states with a single phase-only SLM \cite{36,40,41,42}. We adopt the method \cite{42} to modulate the wavefronts according to the computer-generated holograms that are specifically calculated for maximizing the state fidelity.

\begin{figure}[tbph]
\includegraphics[width=5.5in]{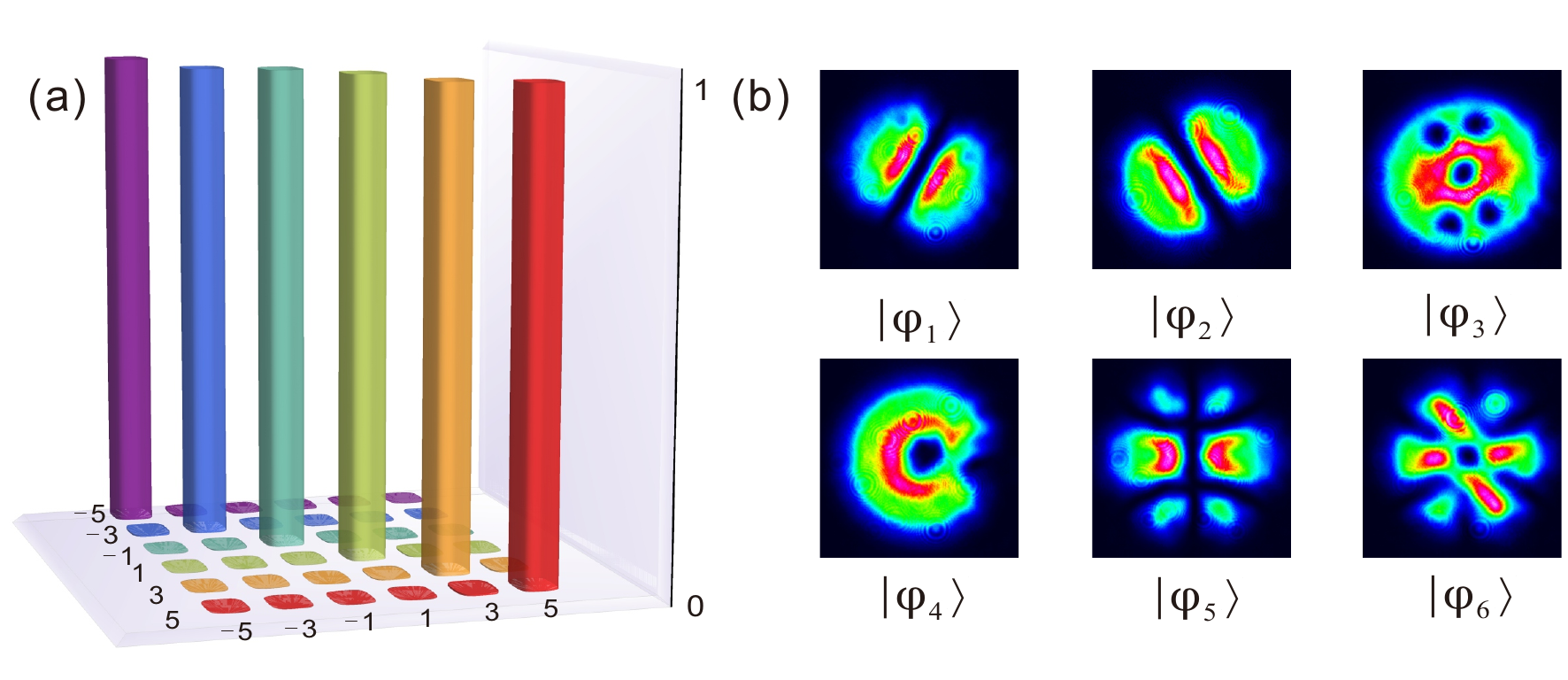}
\caption{Experimental analysis of different OAM photonic qudit states. (a) Cross-talk matrix between the different OAM modes. (b) Experimentally produced intensity profiles of the OAM quantum states used in our experiment.}
\end{figure}

We use two SLMs to realize quantum-state preparation and measurement: SLM1 is responsible for state preparation and SLM2 for state measurement. To implement these functions, we load optimized computer-generated holograms onto SLM1 and SLM2. By combining SLM2 with a single-mode fiber, we can achieve projective measurements. If the incoming photon carries the OAM mode corresponding to a projection of the measurement, the phase of the mode is flattened and the photon will couple to the single-mode fiber. By changing the holograms displayed on SLM2, we can use the same experimental setup to achieve different measurements.

In our experiment, the Gaussian diameter of the beam after the fiber coupler (FC) is about 2500 $\rm \mu m$. After spatial filtering and expansion using two lenses, the Gaussian diameter increases to 4260 $\rm \mu m$. The OAM state of the signal photons is manipulated by SLM1 to prepare the desired state for Alice. For SLM1, we employ optimized holograms to enhance the fidelity of quantum state preparation. The holograms are generated using the method described in \cite{42}, which is carefully designed to maximize the state fidelity. This method can simultaneously control the amplitude and phase of light via computer-generated holograms, and undergoes specialized computations to maximize state fidelity. Then the light subsequently traverses a 4f system to mitigate the Gouy phase-shift effect before reaching SLM2.

\begin{figure}[tbph]
\centering
\includegraphics[width=4.0in]{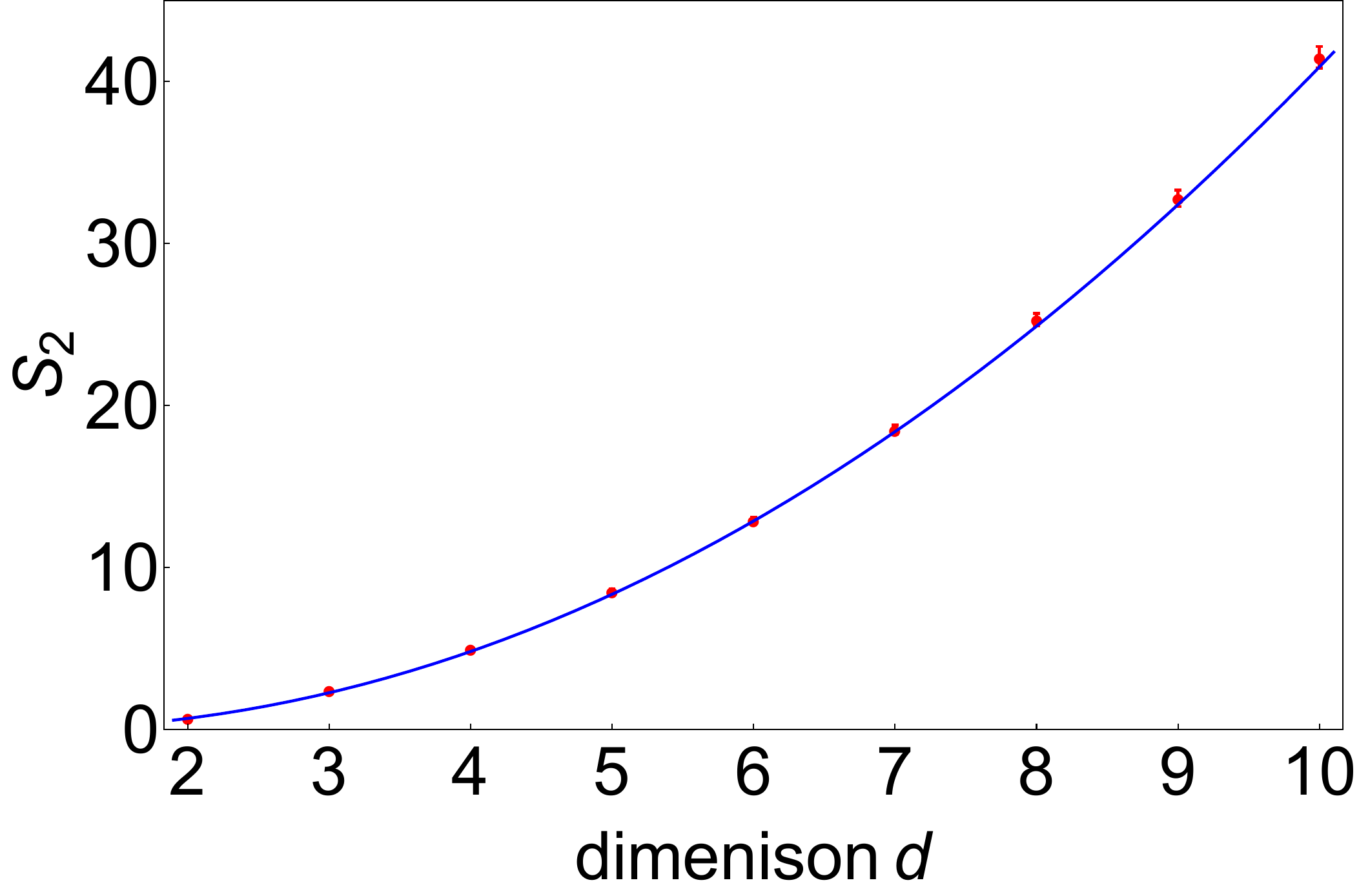}
\caption{Experimental results of certifying high-dimensional quantum states with independent devices. In our experiment, we certify high-dimensional quantum states with dimensions 2-10. We randomly prepare initial states and measure them on different bases. Once the measured values of $S_{2}$ are equal to the theoretical values, we can determine which quantum states the initial states correspond to.}
\end{figure}

For SLM2, two methods are employed to improve the fidelity of measurements. First of all, we use the method described in \cite{42} to produce optimized holograms, which computes phase patterns optimized to maximize the fidelity of measurement. Secondly, in the experiment, we need to align the beam center with the center of the hologram loaded on the SLM2. We find that even a misalignment of approximately 25 $\rm \mu m$ between the hologram center and the optical beam center significantly reduces the fidelity. Conventional manual adjustment of the SLM to align the hologram center with the beam center can introduce substantial errors. Our method generates holograms larger than the screen of SLM and shifts the position of holograms on a computer rather than mechanically moving the SLM device. This approach reduces the crosstalk between different OAM modes and thus improves the fidelity of the measurements. After passing another 4f system, the light is collected by a single-mode fiber and a single-photon detector. We also show the crosstalk matrix for the OAM states used in our experiments, as illustrated in Fig.~3(a). In our experiments, the crosstalk of the OAM states is very small. Figure 3(b) shows some intensity profiles of the high-dimensional OAM quantum states. In our experiment, the average fidelity of the six-dimensional OAM states is 99.0$\%$, we also calculate the similarity parameter of OAM states for up to ten dimensions \cite{Supplemental}.

\begin{figure}[tbph]
\centering
\includegraphics[width=4.0in]{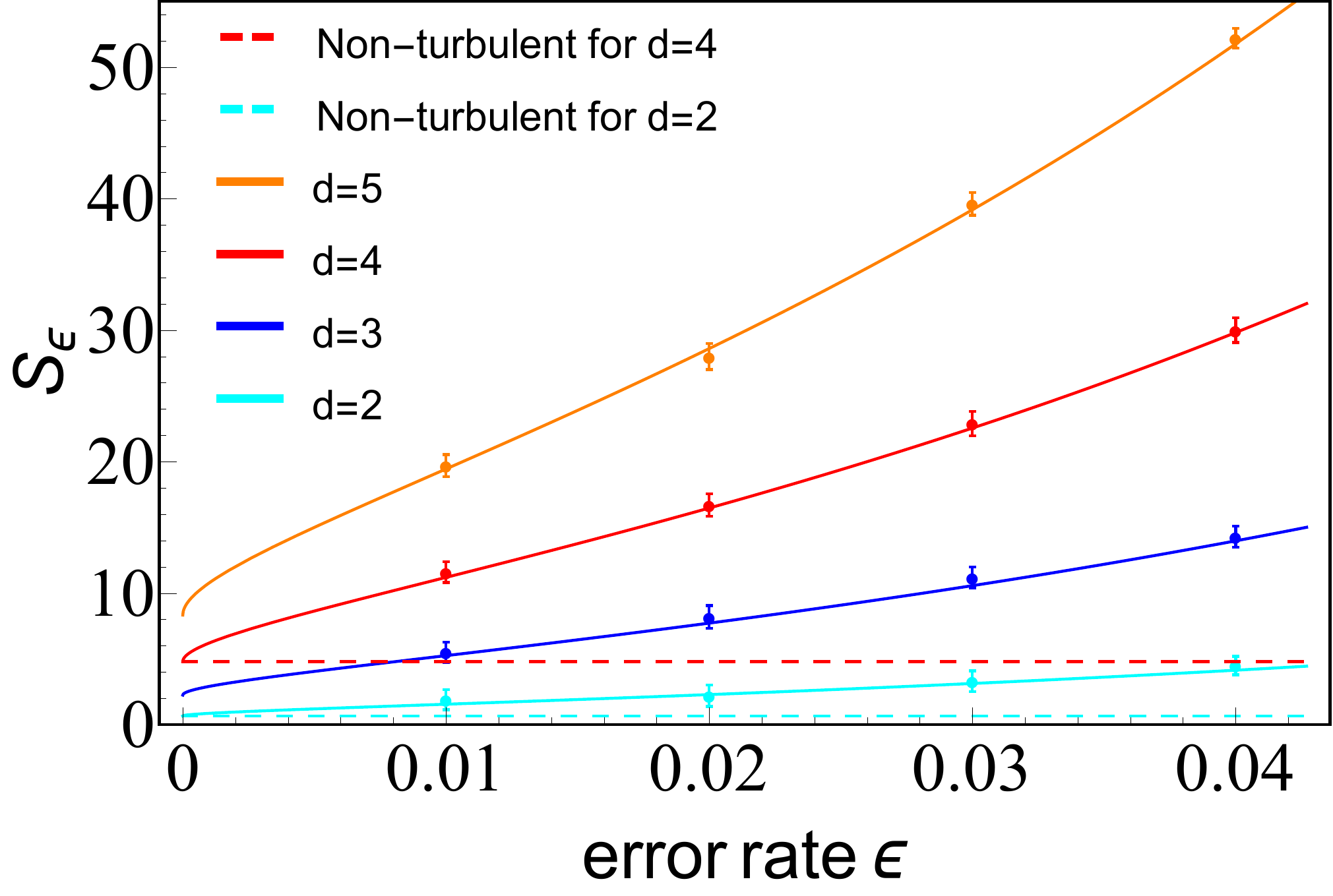}
\caption{Certifying high-dimensional quantum states under the effects of atmospheric turbulence. We investigate the relationship between $S_{\epsilon}$ and the error rate $\epsilon$. Atmospheric turbulent noise affects Bob's error rate $\epsilon$. As the error rate $\epsilon$ increases, the value of $S_{\epsilon}$ gradually increases. The theoretical predictions are in good agreement with the experimental results.}
\end{figure}

The experimental results for the quantum state certification are shown in Fig.\ 4. In our experiment, we certify a family of quantum $t$ designs, which correspond to $t=2$ and $N=d^{2}$ for any $d\geq2$; namely, we need to measure the inconclusive events $S_{2}$. Here, we show the certification of quantum states up to dimension ten. We start the experiment with $d=2$. The value $S_{2}$ of the certification for $d=2$ was found to be $0.69\pm0.0078$. We then repeat the experiment for a higher dimension. We continue this process for dimensions from 3 to 10. When the dimension is 10, we find that $S_{2}$ is $41.5\pm0.6724$. Therefore, according to these experimental results, we can certify the symmetric-informationally-complete quantum states with dimension up to ten. In our work, the statistics of photon counts is assumed to follow a Poisson distribution. The error bars are generated by propagating the Poissonian noise on the measurement outcomes.

We further investigate the quantum state certification under discrimination errors. When Bob's discrimination is subject to an error rate $\epsilon$, based on the observed error rate, all the inconclusive events ($x_{1}<x_{2}$) accumulated across the individual USD tasks are \cite{Armin2020}:
\begin{equation}
S_{\epsilon}\equiv \sum_{x_{1}, x_{2}=1}^{N}(\alpha_{\epsilon}-p^{x_{1},x_{2}}_{\mathrm{usd}})^{2t}.
\end{equation}
where 
\begin{equation}
\alpha_{\epsilon} = \dfrac{1-\epsilon}{(1-2\epsilon)^{2}}[1+2\sqrt{\epsilon(1-\epsilon)}].
\end{equation}

A significant advantage of high-dimensional quantum states is their increased information capacity per quantum system, thus allowing for higher information rates. Over the past decade, significant advancements have been made in the generation and manipulation of high-dimensional quantum states utilizing the OAM of photons. However, a primary challenge persists in ensuring their reliable transmission between remote locations. High-dimensional quantum states that rely on OAM tend to become distorted when subjected to perturbations, and the turbulence in free-space atmospheric links is a striking example. Therefore, we investigate the impact of atmospheric turbulence on high-dimensional OAM states, focusing on how turbulence-induced noise affects the certification of high-dimensional quantum states.

The atmospheric turbulence is simulated by a single phase screen based on the Kolmogorov theory of turbulence \cite{43, 44}. The scintillation strength of the random phase function is given by $W=w_{0}/r_{0}$ \cite{43}, where $w_{0}$ is the radius of the optical beam and $r_{0}$ represents the Fried parameter in the random phase function. Here, $w_{0}$ and $r_{0}$ can be adjusted using SLM2. In our experiment, we can use the magnitude of $W$ to represent the intensity of the turbulent noise generated in the actual experiment \cite{Alpha2014}. As shown in Ref. \cite{43}, negativity decreases gradually with increasing the intensity of the turbulence noise, which proves that this is an effective method. In our experiment, the simulated-turbulence hologram is added to the measurement hologram displayed on SLM2. Namely, to measure OAM states in the presence of turbulent noise, the random phase hologram and the measurement hologram must be displayed simultaneously on SLM2 \cite{Supplemental}. Based on the SLM2, we can achieve the measurement of quantum states, which in turn allows us to determine the optimal success probability. The relation between the optimal success probability and the error rate $\epsilon$ can be found in the supplementary material [see Eq. (S3) there]. Given the optimal success probability, we can calculate the error rate $\epsilon$ according to Eq. (S3). Thus, we can experimentally generate a turbulence hologram with the given error rate $\epsilon$.

In order to show the certification under discrimination errors, we realize the certification under the effects of atmospheric turbulence. Since the atmospheric turbulence affects the error rate $\epsilon$ of Bob's discrimination, one can see that as the error rate $\epsilon$ increases from Fig. 5, the value of $S_{\epsilon}$ also increases. In our experiment, we certify ensembles of high-dimensional quantum states subjected to the influence of atmospheric turbulence, with dimension up to five.

In our experiments, the error rate $\epsilon$ is related to the intensity of the turbulence noise (i.e., turbulence strength). The atmospheric turbulence is simulated by adding random phase fluctuations to the phase hologram displayed on SLM2. In this way, the turbulent strength is described by the scintillation strength $W$ of the random phase fluctuations, i.e., we can use the magnitude of $W$ to represent the strength of the turbulence noise generated in the actual experiment. Obviously, changing the scintillation strength $W$ will influence the optimal success probability $P_{\mathrm{usd}}$. As shown in the supplementary materials, we can experimentally measure the relation between the scintillation strength $W$ (i.e., the intensity of turbulence noise generated in the actual experiment) and the optimal success probability $P_{\mathrm{usd}}$. By combining the measured $P_{\mathrm{usd}}$-$W$ relation with this $P_{\mathrm{usd}}$-$\epsilon$ relation, one obtains the relation between the scintillation strength $W$ and the error rate $\epsilon$, and thus the relation between the turbulence strength and the error rate $\epsilon$.

In conclusion, we have experimentally certified ensembles of quantum states with dimension up to ten based on prepare-and-measure experiments. In our experiment, the preparer and the measurer have no shared randomness. We have certified ensembles of quantum states through a randomized version of unambiguous state discrimination in a SDI manner. We have prepared the quantum states in the high dimensional OAM state with high fidelity. Furthermore, we have investigated the certification of quantum states under the effects of noise. To the best of our knowledge, this work is the first to report an experimental test of certifying ensembles of high-dimensional quantum states with independent devices. Our results offer intriguing prospects for SDI certification of ensembles of quantum states, which has implications in quantum foundations and is of importance in quantum information science and technology.

We are most grateful to Zhao-Di Liu and Rui-Heng Miao for their useful discussion on this manuscript. This work was supported by the National Key Research and Development Program of China (Grant No. 2024YFA1408900), the National Natural Science Foundation of China (62105086, U21A20436), the Hangzhou Joint Fund of the Zhejiang Provincial Natural Science Foundation of China under Grant No. LHZSD24A050001, Scientific Research Foundation for Scholars of HZNU (4085C50221204030), and the Innovation Program for Quantum Science and Technology (Grant No. 2021ZD0301705). F.N. is supported in part by: the Japan Science and Technology Agency (JST) [via the CREST Quantum Frontiers program Grant No. JPMJCR24I2, the Quantum Leap Flagship Program (Q-LEAP), and the Moonshot R$\&$D Grant Number JPMJMS2061].




\end{document}